\documentstyle[secnumtab,fbsedit2]{fbsart11}

\newcommand{\bqn}{\begin{eqnarray}}
\newcommand{\eqn}{\end{eqnarray}}
\newcommand{\beq}{\begin{equation}}
\newcommand{\eeq}{\end{equation}}
\newcommand{\Hh}{{\textstyle\frac 12}}
\newcommand{\Hf}{{\textstyle\frac 14}}
\newcommand{\mbf}[1]{\mbox{\boldmath$#1$}}
\newcommand{\nrmp}{|\mbox{\boldmath $p$}|}

\newcommand{\gve}{\varepsilon}

\title{Light-front versus Bethe-Salpeter forms of two nucleon
  amplitudes\thanks{Dedicated to
    Professor Hartmuth Arenh\"ovel on the occasion of his 60th
    birthday}} \author{S.G. Bondarenko$^1$, V.V.  Burov$^1$, M.
  Beyer$^2$, and S.M. Dorkin$^3$} \institute{$^1$ Bogoliubov
  Laboratory of Theoretical Physics, JINR Dubna,
  141980 Russia\\
  $^2$ Department of Physics,
  University of Rostock, 18051 Rostock, Germany\\
  $^3$ Far Eastern State University of Vladivostok, 690000 Russia}

\runningauthor{S.G. Bondarenko et al.}
\runningtitle{Light-front versus Bethe-Salpeter approaoch}

\begin{document}

\maketitle

\begin{abstract}
  We discuss the relation between the two nucleon Bethe-Salpeter
  amplitude and the light front wave functions. Both approaches
  provide a covariant description for the deuteron bound state and the
  two nucleon scattering state. A comparison is done for the
  spin-orbit functions as well explicit integrals are given on the
  basis of the Nakanishi integral representation method.
\end{abstract}

\section{Introduction}
The description of deuteron properties and reactions involving the
deuteron is reflected in a large part of Arenh\"ovel's work. Relaying
also on his analyzes~\cite{arenhoevel} of experimental data the
discussion of relativistic issues in reactions involving the deuteron
has become more and more important in recent years. There are well
known examples of clear experimental evidence, in particular in the
electromagnetic disintegration of the deuteron~\cite{cam82,sch92}.

Meanwhile two relativistic approaches to the nucleon nucleon system
have received special attention in the last few years. One is based on
the Bethe-Salpeter equation~\cite{sal51} and its various
three-dimensional reductions. Based on quantum field theory the
Bethe-Salpeter formalism is four-dimensional and explicitly Lorentz
covariant.  Interactions (e.g. electromagnetic currents) are
consistently treated via the Mandelstam formalism leading to Feynman
diagrams and the corresponding rules. The second approach considered
here is based on light front dynamics~\cite{dir49}. The state vector
describing the system is expanded in Fock components defined on a
hypershere in the four-dimensional space time. This approach is
intuitively appealing since it is formally close to the
nonrelativistic conception in terms of Hamiltonians, and state vectors
maybe directly interpreted as wave functions.

The equivalence between these field theoretic and light front
approaches has been a subject of recent discussions, see
e.g.~\cite{equiv} and references therein.  A comparison of both
approaches for the deuteron as a two body system clarifies the
structure of the different components of the amplitude. It is also
useful in the context of three particle dynamics where the proper
covariant and/or light front construction of the nucleon amplitude in
terms of three valence quarks (including spin dependence and
configuration mixing) is presently discussed~\cite{bkw98,karm98}.
Although the relation between light front and Bethe-Salpeter
amplitudes for the two nucleon amplitude has been spelled out to some
extend in a recent report~\cite{car98} we provide here some useful
details and additionally discuss the use of the Nakanishi
representation.

To proceed we first present different ways to construct a complete
(covariant) Bethe-Salpeter amplitude, see, e.g. Ref.~\cite{BBBD}. In
particular, we consider the so called direct product and the matrix
representation form.  Besides the spin structure of the wave function
(amplitudes) we also present a comparison of the ``radial'' part of
the amplitude on the basis of the Nakanishi integral
representation~\cite{nakanishi}. This integral representation -- well
known and elaborated for the scalar case, see, e.g.
Ref~\cite{williams}, however, not so frequently used -- allows us to
establish a connection between the different approaches also for the
weight functions (or densities) that has not been done so far and is
relevant for a treatment of the Bethe-Salpeter equation in Minkowski
space.

In the following we present different ways to construct a complete
(covariant) Bethe-Salpeter amplitude.  In this context the so called
direct product representation used in the rest frame of the nucleon
nucleon system using the $\rho$-spin notation is close to the
nonrelativistic coupling scheme and provides states of definite
angular momentum. To construct the covariant basis this form is
transformed into a matrix representation which will then be expressed
in terms of Dirac matrices. A generalization to arbitrary deuteron
momenta finally leads to the covariant representation of the
Bethe-Salpeter amplitude. This will be explained in the next section
along with an explicit construction of the deuteron ($J=1$) and the
$J=0$ nucleon nucleon state.

The construction of the light front form from the Bethe-Salpeter
amplitude will be given in Section~\ref{sect:lfbs}. Presentation of
the light front approach will be kept concise here, since is has been
presented at length in a recent report~\cite{car98}. Again we show the
results for $J=1$ and $J=0$ deuteron and scattering states. Finally,
we present the analysis in terms of the Nakanishi integral
representation.

\section{The Bethe-Salpeter approach to the two nucleon system}

Commonly, two forms are utilized to describe the Bethe-Salpeter wave
functions (amplitudes) known as direct product form and matrix form.
They will be explained in the following.

For convenience, we introduce the $\rho$-spin notation for Dirac
spinors with momentum \mbf{p}, and spin projection $\mu $,
\begin{equation}
U_{\mu}^{\rho}(\mbf{p})=
\left\{\begin{array}{ll} u_{\mu}(\mbf{p}),&\rho=+\\
v_{-\mu}(-\mbf{p}),&\rho=-\end{array}\right.
\end{equation}
The Dirac spinors $u_\mu(\mbf {p})$, $v_\mu(\mbf{p})$
are defined according to Ref. \cite{itzik}, viz.
\begin{eqnarray}
u_{\mu}(\mbf{p}) =
{\cal L}(\mbf{p}) u_{\mu} (\mbf{0}), \quad
v_{\mu}(\mbf{p}) =
{\cal L}(\mbf{p}) v_{\mu} (\mbf{0}).
\label{trans}
\end{eqnarray}
where the boost of a spin-$\Hh$ particle with mass $m$ is given by
\begin{eqnarray}
{\cal L}(\mbf{p}) =
 \frac {m+ p\cdot\gamma \gamma_0}{\sqrt{2E(m+E)}},
\label{lor}
\end{eqnarray}
where the nucleon four momentum is $p=(E,\mbf{p})$, and
$E=\sqrt{\mbf{p}^2+m^2}$.  In the rest frame of the particles the
spinors are given by
\begin{eqnarray}
u_{\mu}(\mbf{0}) =
\left(\begin{array}{c} \chi_{\mu}  \\ 0\ \end{array} \right),  \quad
v_{\mu}(\mbf{0})=
\left(\begin{array}{c}         0   \\ \chi_{-\mu} \ \end{array} \right).
\nonumber
\end{eqnarray}

In the direct product form the basis of the two particle spinor in
the rest frame is represented by
\begin{equation}
{U_{\mu_1}^{\rho_1}}(\mbf{p})\;
 {U_{\mu_2}^{\rho_2}}(-\mbf{p}).
\end{equation}
Which one of the combinations $\rho_1\rho_2$ are actually present in a
particular amplitude depends on the parity and permutation symmetries
required. E.g. since $U^+U^-$ is parity odd, in the deuteron this
combination could only appear for $L=1$. Therefore the appearance of
$P$-states is a typical relativistic effect, because $U^-\rightarrow
0$ for the nonrelativistic limit.
The spin-angular momentum part ${\cal Y}_M^{\alpha}(\mbf{p})$ of the
two nucleon  amplitude is then given by
\begin{eqnarray}
{\cal Y}_M^{\alpha}(\mbf{p})=
i^L \sum \limits_{\mu_1 \mu_2 m_L}\;
\langle L m_L S m_S | J M\rangle \;
\langle\Hh \mu_1 \Hh \mu_2 | S m_S\rangle\;
Y_{L m_L}({\hat{\mbf{p}}})\;
U_{\mu_1}^{\rho_1}(\mbf{p})\;
U_{\mu_2}^{\rho_2}(-\mbf{p}),
\label{Ydecomp}
\end{eqnarray}
where $\langle\cdot|\cdot\rangle$ denotes the Clebsch-Gordan
coefficient, and ${\hat {\mbf{p}}} =
{\mbf{p}}/{|\mbf{p}|}$. The decomposition is according to the quantum
numbers of relative orbital angular momentum $L$, total spin $S$,
total angular momentum $J$  with projection $M$, and $\rho$-spin
$\rho_1$, $\rho_2$, collectively denoted by $\alpha$ \cite{kubis}. The
Bethe-Salpeter amplitude of the deuteron with mass $M_d$ is then
written in the following way (see also ~\cite{honzava})
\begin{eqnarray}
\Phi_{JM}(\stackrel{\circ}{P},p)=\sum \limits_{\alpha} \;
g_{\alpha} (p_0,\nrmp)\;
{\cal Y}_M^{\alpha}(\mbf{p}),
\label{reldp}
\end{eqnarray}
where $\stackrel{\circ}{P}=(M_d,\mbf{0})$.  The radial parts of the
wave function are denoted by $g_{\alpha} (p_0,\nrmp)$.

The matrix representation of the Bethe-Salpeter
amplitude~\cite{nakanishi} is obtained from the above expression
Eq.~(\ref{reldp}) by transposing the spinor of the second particle. In
the rest frame of the system this reads for the basis spinors
\begin{equation}
{U_{\mu_1}^{\rho_1}}(\mbf{p})\;
 {U_{\mu_2}^{\rho_2}}(-\mbf{p})\;
\longrightarrow \;
{U_{\mu_1}^{\rho_1}}(\mbf{p})\;
 {U_{\mu_2}^{\rho_2\top}}(-\mbf{p}),
\label{replace}
\end{equation}
which is now a $4\times 4$ matrix in the two particle spinor space.
The nucleon nucleon Bethe-Salpeter wave function in this basis
is then represented by
\begin{eqnarray}
\Psi_{JM}(P_{(0)},p)=\sum \limits_{\alpha}\;
g_{\alpha} (p_0,\nrmp)\;
\Gamma_M^{\alpha}(\mbf{p})\;C.
\label{reldmatrix}
\end{eqnarray}
where $C$ is the charge conjugation matrix, $C=i\gamma_2\gamma_0$, and
$\Gamma_M^{\alpha}$ is defined as ${\cal Y}_M^{\alpha}$ where the
replacement Eq.~(\ref{replace}) is used.

As an illustration we give an example how to calculate the
spin-angular momentum part of the vertex function for the $^3
S_1^{++}$ state, where we use the spectroscopic notation
$^{2S+1}L^{\rho_1 \rho_2}_{J}$ of Ref.~\cite{kubis}. In this case
\begin{eqnarray}
\lefteqn{\sqrt{4 \pi}\; {\Gamma}^{^3 S_1^{++}}_M( \mbf{p})
 = \sum \limits _{\mu_1 \mu_2}\;
\langle\Hh \mu_1 \Hh \mu_2 |1M\rangle\; u_{\mu_1}( \mbf{p})\;
u_{\mu_2}^\top(-\mbf{p})}&&\nonumber\\
&=&{{{\cal L}}}( \mbf{p})\sum \limits _{\mu_1 \mu_2}
\;\langle \Hh \mu_1 \Hh \mu_2 |1M\rangle
\;\left(\begin{array}{c} \chi_{\mu_1} \\ 0 \end{array} \right)
\;\left(\chi_{\mu_2}^\top\; 0\right )
\;{{{\cal L}}}^\top(-\mbf{p})\nonumber\\
&=&{{{\cal L}}}( \mbf{p})
\;\left( \begin{array}{cc}
\sum \limits_{\mu_1 \mu_2}
\langle\Hh \mu_1 \Hh \mu_2 |1M\rangle
\chi_{\mu_1} \chi^\top_{\mu_2} &0 \\ 0&0 \end{array}\right)
\;{{{\cal L}}}^\top(-\mbf{p})\nonumber\\
&=&{{{\cal L}}}( \mbf{p})
\;\frac {1+\gamma_0}{2} \frac 1{\sqrt2}
\;\left( \begin{array}{cc} 0 & -\mbf{\sigma}\cdot \mbf{\epsilon}_M\\
 \mbf{\sigma}\cdot\mbf{\epsilon}_M &0 \end{array}\right)
\;\left( \begin{array}{cc} 0 & -i \sigma_2\\
-i\sigma_2 &0 \end{array}\right)
\;{{{\cal L}}}^\top(-\mbf{p})\nonumber\\
&=&{{{\cal L}}}( \mbf{p})
\;\frac {1+\gamma_0}{2} \frac 1{\sqrt2}
\;(-\mbf{\gamma}\cdot\mbf{\epsilon}_M)\;{{{\cal L}}}( \mbf{p})
C\nonumber\\
& =&\frac{1}{2E(m+E)}\frac{1}{\sqrt{2}}\;(m+{\gamma\cdot p_1})\;
\frac{1+\gamma_0}{2}\; \gamma\cdot\epsilon_M\;(m-{\gamma\cdot p_2})\;C,
\nonumber
\end{eqnarray}
Here we make use of $\sqrt{2}\sum_{\mu_1 \mu_2} (\Hh \mu_1 \Hh
\mu_2 |1M) \chi_{\mu_1} \chi^T_{\mu_2}=
(\mbf{\sigma}\cdot\mbf{\epsilon}_M) \;({i \sigma_2})$, where  $
\mbf{\epsilon}_M$ is the polarization vector of the spin-1 composite
system with the components in the rest frame given by
\begin{eqnarray}
\mbf{\epsilon}_{+1}=(-1,-i,0)/\sqrt{2}, \quad
\mbf{\epsilon}_{-1}=(1,-i,0)/\sqrt{2}, \quad
\mbf{\epsilon}_{0}=(0,0,1),
\label{vecpol}
\end{eqnarray}
and the four-vector $\epsilon_M = (0,\mbf{\epsilon}_M)$. This
replacement can be done for all Clebsch-Gordan coefficients that in
turn allows us to write the basis in terms of Dirac matrices.

To keep the notation short the $\rho$-spin dependence is taken out of
the matrices and therefore the spin-angular momentum
functions ${\Gamma}^\alpha_M(\mbox {\boldmath$p$})$ are
replaced in the following way
\begin{eqnarray}
{\Gamma}^{\tilde\alpha,\, ++}_M(\mbox {\boldmath$p$}) &=
&\frac{\gamma\cdot p_1 + m}{\sqrt{2E(m+E)}}\;
\frac{1+\gamma_0}{2}\;
{\tilde \Gamma}^{\tilde\alpha}_M(\mbox {\boldmath$p$})\;
\frac{\gamma\cdot p_2 - m}{\sqrt{2E(m+E)}},
\nonumber\\
{\Gamma}^{\tilde\alpha,\, --}_M(\mbox {\boldmath$p$}) &=
&\frac{\gamma\cdot p_2 - m}{\sqrt{2E(m+E)}}\;
\frac{-1+\gamma_0}{2}\;
{\tilde \Gamma}^{\tilde\alpha}_M(\mbox {\boldmath$p$})\;
\frac{\gamma\cdot p_1 + m}{\sqrt{2E(m+E)}},
\nonumber\\
{\Gamma}^{\tilde\alpha,\, +-}_M(\mbox {\boldmath$p$}) &=
&\frac{\gamma\cdot p_1 + m}{\sqrt{2E(m+E)}}\;
\frac{1+\gamma_0}{2}\;
{\tilde \Gamma}^{\tilde\alpha}_M(\mbox {\boldmath$p$})\;
\frac{\gamma\cdot p_1 + m}{\sqrt{2E(m+E)}},
\nonumber\\
{\Gamma}^{\tilde\alpha,\, -+}_M(\mbox {\boldmath$p$}) &=
&\frac{\gamma\cdot p_2 - m}{\sqrt{2E(m+E)}}\;
\frac{1-\gamma_0}{2}\;
{\tilde \Gamma}^{\tilde\alpha}_M(\mbox {\boldmath$p$})\;
\frac{\gamma\cdot p_2 - m}{\sqrt{2E(m+E)}},
\label{gf}
\end{eqnarray}
with $\tilde\alpha=~^{2S+1}L_J$. The matrices ${\tilde
\Gamma}^{\tilde\alpha}$ for $J=0,1$ states are given later in
Tabs.~\ref{tab:1s0} and  \ref{tab:3s1}.

To conclude this  paragraph we give the following useful relations.
The adjoint functions are defined through
\begin{eqnarray}
{\bar {\Gamma}_M^{\alpha}}(\mbox {\boldmath$p$})
=\gamma_0 \;\left[{{\Gamma}_M^{\alpha}}
(\mbox {\boldmath$p$})\right]^{\dagger}\;\gamma_0,
\label{conj}
\end{eqnarray}
and the orthogonality condition is given by
\begin{eqnarray}
\int d^2{\hat{\mbox {\boldmath$p$}}}\
{\rm Tr} \{ {{{{\Gamma}}_M^{\alpha}}^{\dagger}(\mbox {\boldmath$p$})
{\Gamma}_{M^{\prime}}^{\alpha^{\prime}}
(\mbf{p})
\} = \delta_{M {M^{\prime}}}
\delta_{\alpha {\alpha}^{\prime}}}.
\label{ortm}
\end{eqnarray}
In addition, for identical particles the Pauli principle holds, which
reads
\begin{equation}
\Psi_{JM}(\stackrel{\circ}{P},p)=-P_{12}\Psi_{JM}(\stackrel{\circ}{P},p)
=(-1)^{I+1}C\left[\Psi_{JM}(\stackrel{\circ}{P},-p)\right]^\top C.
\end{equation}
where $I$ denotes the channel isospin.
This induces a definite  transformation property of the
radial functions $g_{\alpha} (p_0,\nrmp)$
on replacing $p_0 \rightarrow -p_0$, which is even or odd,
depending on $\alpha $. Also,
since the $P^{\rho_1\rho_2}$ amplitudes
do not have a definite symmetry we use instead
\begin{eqnarray}
\Gamma^{P^e}_M &=&
\frac {1}{\sqrt 2}(\Gamma^{P,+-}_M
+ \Gamma^{P,-+}_M),\nonumber \\
\Gamma^{P^o}_M &=&
\frac {1}{\sqrt 2}(\Gamma^{P,+-}_M - \Gamma^{P,-+}_M).
\label{Geven}
\end{eqnarray}
These functions have definite even(e) or odd(o) $\rho$-parity, which
allows us to define a definite symmetry behavior under particle
exchange.

We now discuss the $^1S_0$
channel and the deuteron channel in some detail.

\subsection{The $^1S_0$ channel}

For the two nucleon system in the $J=0$ state the relativistic wave
function consists of four states, i.e.  $^1S_0^{++}$, $^1S_0^{--}$,
$^3P_0^{e}$, $^3P_0^{o}$, labeled by $1,\dots,4$ in the following.
The Dirac matrix representation of the spin structures are shown in
Table~\ref{tab:1s0}.
\begin{table}[h]
\caption{\label{tab:1s0} Spin angular momentum parts
$\tilde \Gamma_0^{\tilde\alpha}$
for the $J=0$ channel}
\[
\begin{array}{cc}
\hline\hline
\tilde\alpha&{\sqrt{8\pi} \;\;\tilde \Gamma}_0^{\tilde\alpha}\\[1ex]
\hline
^1S_0&-\gamma_5\\[1ex]
^3P_0&\nrmp^{-1} ({\gamma\cdot p_1}-{\gamma\cdot p_2}) \gamma_5\\
\hline\hline
\end{array}
\]
\end{table}
Note the formally covariant relation for $\nrmp$, and also for $p_0$
and $E$, used in the following,
\begin{equation}
p_0=\frac{P\cdot p}{M}, \quad E=\sqrt{\frac {(P\cdot p)^2}{M^2}-p^2+m^2},
\quad \nrmp= \sqrt{\frac {(P\cdot p)^2}{M^2}-p^2}.
\nonumber
\end{equation}

Eq.~(\ref{gf}) along with Table~\ref{tab:1s0} may now be used as a
guideline to construct covariant expressions for the $J=0$ nucleon
nucleon Bethe-Salpeter amplitude. This will be achieved by allowing
the momenta involved to be off-shell. Introducing then four Lorentz
invariant functions $h_i(P\cdot p,p^2)$ this amplitude is given by
\begin{eqnarray}
\Psi_{00}(P,p) &=
&h_1\gamma_5 +
h_2\frac {1}{m} (\gamma\cdot {p}_1\gamma_5 +\gamma_5 \gamma\cdot {p}_2)
\nonumber\\&&
+h_3\left(\frac{\gamma\cdot {p}_1-m}{m}
\gamma_5 -\gamma_5 \frac {\gamma\cdot {p}_2+m}{m}\right)\nonumber\\
&&+h_4 \frac{\gamma\cdot {p}_1-m}{m} \gamma_5 \frac {\gamma\cdot
{p}_2+m}{m}
\label{covarj0}
\end{eqnarray}
The connection between the invariant functions $h_i(P\cdot p,p^2)$ and the
functions $g_i (p_0,|{\bf p}|)$ given before is achieved by expanding
the Dirac matrices appearing in Eq.~(\ref{covarj0}) into the
$\Gamma^\alpha$.  The resulting
relation is
\begin{eqnarray}
h_1 &= &- \sqrt{2}D_1\; ( g_1+ g_2)
- \mu p_0 \nrmp^{-1}\; g_3 - 4m\nrmp^{-1}D_0 \; g_4 \nonumber\\
h_2 &=&\Hf m \nrmp^{-1} \; g_3 \nonumber\\
h_3 &=& 8 a_0m^2 (g_1+g_2) - \Hh\mu p_0 \nrmp^{-1}\;g_3
- 8a_0m\nrmp^{-1}\gve  (m-E) \; g_4 \nonumber\\
h_4 &=& -4a_0\sqrt{2}m^2(g_1+g_2) + 8a_0m^3\nrmp^{-1}\;g_4
\end{eqnarray}
where $a_0=1/(16ME)$, $\gve =2m+E$, $\mu=m/M$, $M=\sqrt{(p_1+p_2)^2}$,
and
\begin{eqnarray}
D_0&=&a_0(4p_0^2+16m^2-12E^2-M^2),\\
D_1&=&a_0(-M^2/4+p_0^2-E^2+16m^2+ME).
\end{eqnarray}
Note, that only $h_2$ and $g_3$ are odd with respect to
$p_0\rightarrow -p_0$, and all other functions are even.

\subsection{$^3S_1- ^3D_1$ channel}
In the deuteron channel the relativistic wave function consists of
eight states, i.e. $^3S_1^{++}$, $^3S_1^{--}$,$^3D_1^{++}$,
$^3D_1^{--}$, $^3P_1^{e}$, $^3P_1^{o}$, $^1P_1^{e}$, $^1P_1^{o}$,
labeled by $1,\dots,8$ in the following. There Dirac matrix
representation of the spin structures ${\tilde
  \Gamma}^{\tilde\alpha}_M$ is shown in Table~\ref{tab:3s1}.
\begin{table}[h]
\caption{\label{tab:3s1} Spin angular momentum parts
$\tilde \Gamma_M^{\tilde\alpha}$
for the deuteron channel}
\[
\begin{array}{cc}
\hline\hline
\tilde\alpha&{\sqrt{8\pi}\;\;\tilde \Gamma}_M^{\tilde\alpha}\\[1ex]
\hline
^3S_1&{\gamma\cdot  \epsilon_M}\\
^3D_1& -\frac{1}{\sqrt{2}}
\left[ {\gamma\cdot  \epsilon_M}+\frac{3}{2}
({\gamma\cdot  p_1}-{\gamma\cdot
p_2})p\cdot\epsilon_M\nrmp^{-2}\right]\\
^3P_1& \sqrt{\frac{3}{2}}
\left[ \frac{1}{2} {\gamma\cdot  \epsilon_M}
({\gamma\cdot  p_1}-{\gamma\cdot  p_2})
-p\cdot\epsilon_M \right]\nrmp^{-1}\\
^1P_1&\sqrt{3} p\cdot\epsilon_M\nrmp^{-1}\\
\hline\hline
\end{array}
\]
\end{table}

Again, generalizing the Dirac representation it is possible to achieve
a covariant form of the Bethe-Salpeter amplitude with eight Lorentz
invariant functions $h_i(P\cdot p,p^2)$,
\begin{eqnarray}
\Psi_{1M}(P,p) &=
&h_1 \gamma\cdot  {\epsilon}_M
+h_2 \frac{p \cdot\epsilon_M}{m} \nonumber\\
&&+h_3 \left (\frac {\gamma\cdot  p_1-m}{m} \gamma\cdot  {\epsilon}_M +
\gamma\cdot {\epsilon}_M \frac{\gamma\cdot  p_2+m}{m}\right)
\nonumber\\
&&+h_4 \left(\frac {\gamma\cdot  p_1 + \gamma\cdot  p_2}{m}\right)
\frac {p\cdot \epsilon_M}{m}\nonumber\\&&+
h_5 \left(\frac {\gamma\cdot  p_1-m}{m} \gamma\cdot  {\epsilon}_M -
\gamma\cdot {\epsilon}_M \frac{\gamma\cdot  p_2+m}{m}\right)
\nonumber\\
&&+h_6 \left(\frac {\gamma\cdot  p_1 - \gamma\cdot  p_2-2m}{m}\right)
\frac {p\cdot \epsilon_M}{m}
\nonumber \\
&&+\frac {\gamma\cdot  p_1-m}{m}\left(h_7 {\gamma\cdot  \epsilon_M}
+h_8 \frac{p\cdot \epsilon_M}{m} \right )\frac{\gamma\cdot  p_2+m}{m}
\label{covar}
\end{eqnarray}
For the deuteron, the functions $h_i(P\cdot p,p^2)$ and $g_i(p_0,\nrmp)$ are
connected via
\begin{eqnarray}
h_1 &= &D^+_1\; (g_3-\sqrt{2}g_1)
        + D^-_1\; (g_4-\sqrt{2}g_2)
        + \Hh\sqrt{6}\mu p_0\nrmp^{-1} \;g_5
        + \sqrt{6}m D_0 \nrmp^{-1} \;g_6               \nonumber \\
h_2 &= &\sqrt{2}(D^-_2\;g_1 + D_2^+\;g_2)
        - D_3^+\;g_3
        - D_3^-\;g_4 \nonumber\\
      &&-\Hh\sqrt{6} \mu p_0\nrmp^{-1}\;g_5
        -\sqrt{6} D_4 m\nrmp^{-1} \;g_6
        +\sqrt{3} m^2 \nrmp^{-1} E^{-1}\;g_7\nonumber \\
h_3 &= &-\Hf\sqrt{6}m \nrmp^{-1}\; g_5\nonumber \\
h_4 &= &8a_1\sqrt{2}mp_0\;(g_1-g_2)
        +8a_2\gve \;mp_0(g_3-g_4)\nonumber \\&&
        -16a_0\sqrt{3}m^2 \nrmp^{-1}\;(p_0g_7-Eg_8)\nonumber \\
h_5 &= &16a_0m^2\;[g_4+g_3-\sqrt{2}\,(g_1+g_2)]\nonumber \\&&
        +8a_0\sqrt{6}m \nrmp^{-1} \;[p_0E\;g_5+
        (2m^2-E^2)\;g_6]   \nonumber \\
h_6 &= &4a_1\sqrt{2}m[D_6^-\;g_1 + D_6^+\; g_2]
        - 4 m^2 \nrmp^{-2} [D_5^+ \;g_3 + D_5^- \;g_4]\nonumber \\&&
        -16a_0\sqrt{6}m^2 \nrmp^{-1}\;(mg_6-M_dg_7)\nonumber \\
h_7 &=& 4a_0 m^2\;[\sqrt{2}(g_1+g_2)-(g_3+g_4)]
        -4a_0\sqrt{6}m^3 \nrmp^{-1}\; g_6 \nonumber \\
h_8 &=& 4a_0m^3 \nrmp^{-2}\;
        [\sqrt{2}(m-E)(g_1+g_2)-(2E+m)(g_3+g_4)]\nonumber \\&&
        +4a_0\sqrt{6}m^3 \nrmp^{-1} \;g_6
\end{eqnarray}
with $a_0$, $\gve $, $\mu$, $D_0$ given above, $a_1=a_0 m/(m+E)$,
$a_2=a_1/(m-E)$, and the dimensionless functions
\begin{eqnarray}
D^\pm_1&=&a_0(4p_0^2+16m^2-M_d^2-4E^2\pm 4M_dE)\\
D^\pm_2&=&a_1(16m^2+16mE+4E^2+M_d^2-4p_0^2\pm 4M_d\gve )\\
D^\pm_3&=&a_2[-12mE^2+2M_d^2E-8p_0^2E+16m^3+mM_d^2-4mp_0^2+8E^3\nonumber\\
            &&  \pm(16m^2M_d+ 4mM_dE- 8E^2M_d)]\\
D_4&=&a_0(16m^2-4E^2-p_0^2+m^2)\\
D^\pm_5&=&a_0(-2E^2+4m^2+4mE\pm \gve M_d)\\
D_6^\pm&=&a_0(2\gve \pm M_d)
\end{eqnarray}
Note now, that $h_3$, $h_4$ and $g_5$, $g_8$ are odd, and all other
functions are even under $p_0\rightarrow -p_0$.

\section{Construction of
the light-front wave function of two nucleon system
from the Bethe-Salpeter amplitude}
\label{sect:lfbs}

We now compare the above given covariant amplitudes of the
Bethe-Salpeter approach to the covariant light front form.  The state
vector defining the light-front plane is denoted by $\omega$, where
$\omega=(1,0,0,-1)$ leads to the standard light front formulation
defined on the frame $t+z=0$.  The formal relation between the
light-front wave functions $\Phi(k_1,k_2,p,\omega \tau )$, depending
on the on-shell momenta $k_1$, $k_2$, and $p=k_1+k_2-\omega\tau$, and
the Bethe-Salpeter amplitude $\Psi (p_1, p_2)$, where $p_1$ and $p_2$
are off-shell momenta has been given in Ref.~\cite{car98},
\begin{equation} \label{bs7}\Phi
(k_1,k_2,p,\omega \tau )=
\frac{k_1\cdot\omega \,k_2\cdot\omega}{\pi \omega\cdot p
\sqrt{m}}\int_{-\infty }^{+\infty }\Psi
(k_1-\omega\tau/2+\omega\beta, k_2-\omega\tau/2-\omega\beta )\,d\beta.
\end{equation}
In the theory on the null plane the integration of Eq.~(\ref{bs7})
corresponds to an integration over $dk^{-}$.  Since $k_1$ and $k_2$
are on the mass shell it is possible to use the Dirac equation after
making the replacement of the arguments indicated in Eq.~(\ref{bs7}).
This will be done explicitly for the $J=0$ and the deuteron channel in
the following.

\subsection{$^1S_0$ channel}

Using the Dirac equations $\bar u(k_1)(\gamma\cdot k_1 -m) =0$, and
$(\gamma\cdot k_2 +m) C u(k_2)^\top=0$ one obtains the following form
of the light front wave function from the Bethe-Salpeter amplitude
using Eq.~(\ref{bs7})
\begin{equation}
\Psi_{00}\rightarrow
H^{(0)}_1  \gamma_5 +
2  H^{(1)}_2 \frac {\beta \gamma\cdot  {\omega}}{m\omega\cdot
P}\gamma_5,
\label{1s0lf}
\end{equation}
The functions $H_1(s,x)$ and $H_2(s,x)$, depending now on
$x=\omega\cdot k_1/\omega\cdot P$ and $s=(k_1+k_2)^2=4(q^2+m^2)$ are
obtained from the functions $h_i(P\cdot p,p^2)$ through the remaining
integrals over $\beta$ implied in Eq.~(\ref{bs7}),
\begin{eqnarray}
H^{(0)}_i(s,x)&=& N \int {h_i((1-2x)(s-M^2)+\beta \omega\cdot P,
-s/4+m^2+(2x-1)\beta)\, \omega\cdot P  d\beta} \nonumber\\
&\equiv&N \int {\tilde h_i (s,x,\beta^{\prime})\,
d\beta^{\prime} },\nonumber\\
H^{(k)}_i(s,x)&\equiv&N \int { \tilde h_i(s,x,\beta^{\prime}) \,
(\beta^{\prime})^k d\beta^{\prime} }
\label{eqn:H}
\end{eqnarray}
where the variable $\beta^{\prime}=\beta \omega\cdot P$ has been
introduced, and $N=x(1-x)$,
$1-x=\omega\cdot k_2/\omega\cdot P$.
The functions $h_3$ and $h_4$ do not
contribute. Instead of the four structures appearing in the
Bethe-Salpeter wave function, the light front function consists of
only two. Note, that the second term in parenthesis is defined by the
pure relativistic component of the Bethe-Salpeter amplitude.

\subsection{$^3S_1$-$^3D_1$ case}

In the deuteron case, starting from formula Eq.~(\ref{bs7}), replacing
the momenta $p_i$, and applying the Dirac equation we arrive at
\begin{eqnarray}
\Psi_{1M}&\rightarrow &
H^{(0)}_1 \gamma\cdot {\epsilon_M} +H^{(0)}_2 \frac {k\cdot \epsilon}{m}
+[H^{(1)}_2+2H^{(1)}_5]
\frac{\omega\cdot \epsilon}{m\omega\cdot P}  \nonumber \\
&&+2H^{(1)}_6  \frac{k\cdot\epsilon \gamma\cdot  {\omega}}{m^2 
\omega\cdot P}
+2 H^{(1)}_3 \frac { \gamma\cdot  {\epsilon} \gamma\cdot  {\omega}
-\gamma\cdot  {\omega} \gamma\cdot  {\epsilon} }{\omega\cdot P}
\nonumber\\&&
+[2H^{(2)}_6+2H^{(2)}_7] \frac { \omega\cdot \epsilon \gamma\cdot
{\omega} }
{m^2(\omega\cdot P)^2},
\label{bsd5}
\end{eqnarray}
where $H_i^{(k)}$ are defined in eq.~(\ref{eqn:H}).  In this case the
functions $h_4$ and $h_8$ do not contribute.  The expression
$(\gamma\cdot {\epsilon}\gamma\cdot {\omega}-\gamma\cdot
{\omega}\gamma\cdot {\epsilon})$ at the term $H_5$ given in
Eq.~(\ref{bsd5}) can be transformed to a different one to compare
directly to the light front form given in Ref.~\cite{car98}.  Using in
addition the on shellness of the momenta $k_1$ and $k_2$ the resulting
form is
\begin{eqnarray}
\label{bsd7}
\bar u_1(\gamma\cdot {\epsilon}\gamma\cdot {\omega}
-\gamma\cdot {\omega}\gamma\cdot {\epsilon})C\bar u_2^\top &=&
\frac{4}{s}\bar u_1[-i\gamma_5e_{\mu\nu\rho\gamma}
\epsilon_{\mu}k_{1\nu}k_{2\rho}\omega_{\gamma}\nonumber\\&&
+k\cdot\epsilon\;\omega\cdot P-m\;\gamma\cdot {\epsilon}  \;\omega\cdot P
\nonumber\\
&&-\frac{1}{2}(s-M^2)(x-\frac{1}{2})\omega\cdot\epsilon
\nonumber\\&&
+\frac{1}{2}m(s-M^2)\,\frac{\gamma\cdot {\omega} \;\omega\cdot\epsilon}
{\omega\cdot P}] C\bar u_2^\top
\end{eqnarray}
The final form of light front wave function then is
\begin{eqnarray}
\Psi_{1M} &\rightarrow &
H_1^{\prime} \gamma\cdot {\epsilon}_M +H_2^{\prime} \frac {k\cdot
\epsilon}{m}
+H_3^{\prime} \frac{\omega\cdot \epsilon}{m\,\omega\cdot P)}
+H_4^{\prime}  \frac{k\cdot \epsilon\; \gamma\cdot  {\omega}}
{m^2 \omega\cdot P}  \nonumber \\&&
+H_5^{\prime} i \gamma_5 e_{\mu \nu \rho
\sigma}\epsilon_{\mu}{k_1}_{\nu}
{k_2}_{\rho}{\omega}_{\sigma}
+H_6^{\prime} \frac { \omega\cdot \epsilon\;
\gamma\cdot {\omega} }{m^2(\omega\cdot P)^2},
\end{eqnarray}
with the functions
\begin{eqnarray}
H_1^{\prime} &=& H^{(0)}_1-\frac{4}{s}2H^{(1)}_3, \nonumber \\
H_2^{\prime} &=& H^{(0)}_2+\frac{4}{s}2H^{(1)}_3, \nonumber \\
H_3^{\prime} &=& [H^{(1)}_2+2H^{(1)}_5]
-\frac{ (s-M^2)}{s}(2x-1) 2H^{(1)}_3, \nonumber \\
H_4^{\prime} &=& 2H^{(1)}_6, \nonumber \\
H_5^{\prime} &=& \frac{4}{ms}2H^{(1)}_3, \nonumber \\
H_6^{\prime} &=& [2H^{(2)}_6+2H^{(2)}_7]
+2 \frac{s-M^2}{s}m^2 2H^{(1)}_3.
\label{eqn:conn}
\end{eqnarray}
Provided the invariant functions $h_i$ are given from a solution of
the Bethe-Salpeter equation the above relations allow us to directly
calculate the corresponding light front components of the wave
functions.

Thus, the projection of the Bethe-Salpeter amplitudes to the light
front reduces the number of independent functions to six instead of
eight for the $^3S_1-^3D_1$ channel and to two instead of four for the
$^1S_0$ channel. The reduction is because the nucleon momenta $k_1$
and $k_2$ are on-mass-shell in the light front formalism.  The result
is based on the application of the Dirac equation and the use of the
covariant form. Any other representation (e.g. spin orbital momentum
basis) also leads to a reduction of the number of amplitudes for the
two nucleon wave function that is however less transparent. For an
early consideration compare, e.g.  Ref.~\cite{GT60}.

\section{Integral representation method}
A deeper insight into the connection of Bethe-Salpeter amplitudes and
light front wave functions will be provided within the integral
representation proposed by Nakanishi~\cite{nakanishi}. This method
has recently been fruitfully applied to solve the Bethe-Salpeter
equation both in ladder approximation and beyond within scalar
theories~\cite{williams}. In this framework the following ansatz for
radial Bethe-Salpeter amplitudes of orbital momentum $\ell$ has been
proposed,
\begin{equation}
\phi_\ell(P\cdot p,p^2)=\int_0^\infty
d\alpha\;\int_{-1}^{+1} dz \;\frac{g_\ell(\alpha,z)}
{(\alpha+\kappa^2-p^2-z\,P\cdot p-i\epsilon)^n},
\label{nak1}
\end{equation}
where $g_\ell(\alpha,z)$ are the densities or weight functions,
$\kappa=m^2-M_d^2/4$ and the integer $n \geq 2$. The weight functions
$g_\ell(\alpha,z)$ that are continuous in $\alpha$ vanish at the
boundary points $z=\pm 1$.  The form eq.~(\ref{nak1}) opens the
possibility to solve the Bethe-Salpeter amplitude in the whole
Minkowski space while commonly solutions are restricted to the
Euclidean space only. In fact the densities could be considered as the
main object of the Bethe-Salpeter theory, because knowing them allows
one to calculate all relevant amplitudes.

For the realistic deuteron we need to expand to Nakanishi form to the
spinor case, which has not been done so far. The key point to do so is
choosing the proper spin-angular momentum functions and perform the
integration over angles in the Bethe-Salpeter equation.  The choice of
the covariant form of the amplitude allows us to establish a system of
equations for the densities $g_{ij}(\alpha,z)$, suggesting the
following general form for the radial functions $h_i(P\cdot p,p^2)$ (even in
$P\cdot p$)
\begin{eqnarray}
h_i(P\cdot p,p^2)&=&\int_0^\infty
d\alpha\;\int_{-1}^{+1} dz\; \left\{\frac{g_{i1}(\alpha,z)}
{(\alpha+\kappa^2-p^2-z\,P\cdot p)^n}\right.\nonumber\\&&
\qquad\qquad\qquad
+\frac{g_{i2}(\alpha,z)\;p^2}
{(\alpha+\kappa^2-p^2-z\,P\cdot p)^{n+1}} \\ \nonumber&&
\qquad\qquad\qquad
\left.+\frac{g_{i3}(\alpha,z)\;
(P\cdot p)^2}{(\alpha+\kappa^2-p^2-z\,P\cdot p)^{n+2}}\right\}.
\label{nak2}
\end{eqnarray}
For the functions that are odd in $P\cdot p$ the whole integrand is
multiplied by factor $P\cdot p$.  Although now the number of densities is
larger the total number of {\em independent} functions is still eight.
The form given in eq.~(\ref{nak2}) is valid only for the deuteron
case. The continuum amplitudes of the $^1S_0$ state, e.g., require a
different form.

The basic point now is that using this integral represention allows us
to perform the integration over $\beta^{\prime}$ in the expressions of
eq.~(\ref{bsd5}).  Substituting the arguments of the functions $h_i$
into the integral representation eq.~(\ref{nak2}) leads to a
denominator of the form
\begin{equation}
{\cal D}^k(\alpha,z;x,s,\beta')=
(\alpha+\frac{s}{4}(1+(2x+1)z)-\beta^{\prime} (2x-1+z)-i\epsilon)^k
\nonumber
\end{equation}
Then, using the identity for an analytic function $F(z)$
\begin{equation}
\int_{-1}^{+1} dz\, \int_0^\infty d\beta^{\prime}\, \frac{F(z)}
{{\cal D}^k(\alpha,z;x,s,\beta')}
=\frac{i\pi}{(k-1)}\frac{F(1-2x)}
{(\alpha+sx(1-x))^{k-1}}
\end{equation}
allows us to express the radial amplitudes in terms of
Nakanishi densities. E.g. $H_5(s,x)$ reads
\begin{equation}
H_5(s,x)=\frac{x(1-x)}{s} \int d\alpha
\left\{ \frac {g_{51}(\alpha,1-2x)}{(\alpha+sx(1-x))}+
\frac{g_{52}(\alpha,1-2x)sx(1-x) }{(\alpha+sx(1-x))^2}\right\}
\end{equation}

Note that the dependence of the amplitude on the light front argument
$x$ is fully determined by the dependence of the density on the
variable $z=1-2x$, which has also been noted in Ref.~\cite{car98} for
the Wick-Cutkosky model.  

This fully completes the connection between the Bethe-Salpeter
amplitude and the light front form. Evaluation of the Nakanishi
integrals does not lead to cancelations of functions. Although some
functions are cancelled for reasons given above all spin orbital
momentum functions (or all densities) in principle contribute to the
light front wave functions.

Once the Bethe-Salpeter
amplitudes are given (or the Nakanishi densities) the
light front wave function can explicitly be calculated. The reduction
of the number of amplitudes is due to $k_1$ and $k_2$ being on-shell
in the light front form. The Nakanishi spectral densities of the
Bethe-Salpeter amplitudes lead directly to the light front wave
function.

\section{Conclusion}
Even more than 60 years after its discovery the deuteron is still an
object of intense research. In recent years the focus on relativistic
aspects has increased, in particular in the context of $ed$ scattering
and relativistic $pd$ reactions. Meanwhile a relativistic description
of the deuteron (i.e. the nucleon nucleon system) has achieved
considerable progress. Two successful relativistic approaches are
based on the Bethe-Salpeter equation or the light front dynamics. In
this paper we provide a detailed comparison of the different
approaches on the basis of the spin-orbital amplitudes and the radial
dependence on the basis of the Nakanishi integral representation. In
this context the $P$ waves of the deuteron play an important role. To
reach this conclusion the covariant form has been given in terms of
the partial wave representation using the $\rho$-spin notation.

We would like to stress that the two relativistic approaches have
shown qualitatively similar results in the description of the
electrodisintegration near threshold. The functions $f_5$ and $g_2$
(notation of Ref.~\cite{car98}) may be related to the pair current in
the light front approach whereas the functions $h_5$ and $h_2$ play
this role in the Bethe-Salpeter approach. The results presented here
allow us to specify this relation on a more fundamental level.

\end{document}